\documentclass{article}
\usepackage[utf8]{inputenc}
\usepackage{mathtools}
\usepackage{authblk}
\usepackage{parskip}
\usepackage{multirow}
\usepackage[toc,page]{appendix}
\usepackage{comment}
\usepackage[sorting=none]{biblatex}
\usepackage{graphicx}
\usepackage{epsfig}
\usepackage{amsmath}
\usepackage{subcaption}
\graphicspath{ {./image/} }
\usepackage{amssymb}
\usepackage{amsfonts}
\usepackage{float}
\numberwithin{equation}{section}
\usepackage{hyperref}
\hypersetup{
    colorlinks=true,
    linkcolor=blue,
    filecolor=magenta,      
    urlcolor=cyan,
    pdftitle={Overleaf Example},
    pdfpagemode=FullScreen,
    }
\usepackage{geometry}
\begin{document}
\date{}
\title{Quantum Correction to Hawking Temperature and Thermodynamics of Schwarzschild AdS black holes}
\author[1]{Prosenjit Paul\thanks{prosenjitpaul629@gmail.com}}
\affil[1]{Indian Institute Of Engineering Science and Technology (IIEST), Shibpur-711103, WB, India}
\maketitle

\begin{center}
\rule{17cm}{0.4mm}
\end{center}

\begin{abstract}
 We investigate quantum corrections to $4D$ Schwarzschild Anti de Sitter black hole using an effective field theory approach. We compute the quantum corrected Schwarzschild AdS spacetime upto order $\mathcal{G}^2$. The Hawking temperature of Schwarzschild AdS black hole is computed upto order $\mathcal{O}{(G^{2})}$  when the emitted object is massless and massive. We use the expression of the modified horizon radius to compute the quantum corrected entropy upto order $\mathcal{O}{(G^{2})}$. Finally, we study quantum corrected thermodynamic parameters specific heat, free energy and pressure.
\end{abstract}
\newpage

\newpage
\begin{center}
\rule{17cm}{0.4mm}
\end{center}
\tableofcontents
\begin{center}
\rule{17cm}{0.4mm}
\end{center}

\section{Introduction}
Black hole are fascinating and elegant objects. They are described very accurately by a small number of macroscopic parameters (e.g., mass, angular momentum, and charge). The surprising discovery that black holes behave as thermodynamic objects has radically affected our understanding of general relativity and its relationship to quantum field theory. In the early 1970s,
Bekenstein \cite{bekenstein2020black,bekenstein2020black1} showed that black holes entropy proportional to area of the event horizon. In 1974, Hawking’s original discovery \cite{hawking1974black,hawking1975black} has showed that black holes can emit thermal radiation at a temperature inversely proportional to their mass. In particular there have been detailed studies \cite{PhysRevD.13.198,PhysRevD.14.3260,PhysRevLett.85.5042,PhysRevLett.85.499,PhysRevD.64.044006} of Hawking temperature and radiation. Particle emission from a non-rotating and rotating black hole has calculated by Page in refs. \cite{PhysRevD.13.198,PhysRevD.14.3260} In 1999, Wilczek and Parikh
showed that Hawking radiation can be viewed as a tunnelling process \cite{PhysRevLett.85.5042}, in which particles pass through the contracting horizon of the black hole, lending formal justification to the most intuitive picture we have of black hole evaporation.

In general, any correction to the black hole entropy and temperature leads to a correction  of the black hole thermodynamics. As the black hole entropy is obtained using area of the event horizon and it is natural to expect that the quantum corrections modify the black hole metric as to reproduce the corrected entropy. Using Effective field theory one can obtain the quantum corrected metric. The quantum corrected metric of Schwarzschild black holes has been obtained recently \cite{calmet2019quantum} using Effective field theory. The effects of quantum correction on the Hawking temperature, thermodynamics and greybody factors of Schwarzschild black holes studied in \cite{liu2022effect}. Quantum correction on Reissner-Nordstrom black holes using Effective field theory approach studied in Refs. \cite{donoghue2015covariant,pourhassan2022quantum}. 

The main aim of the original paper \cite{calmet2019quantum}  is to describe the metric inside and outside of a static and spherical star, taking into account quantum effects. However, the authors also proposed that the exterior metric may provide a good model of quantum black holes. The aim of this paper is to consider quantum correction to Hawking temperature (for massless \& massive particles) and thermodynamics of a static and spherical Schwarzschild AdS black hole. In section \ref{sec:2}, we introduce the effective field theory approach to quantum gravity and obtain the solution of the equations of motion that represents the quantum corrected metric of a Schwarzschild AdS black hole up to order $\mathcal{O}(G^{2})$. In section \ref{sec:3}, we compute quantum corrected Hawking temperature of Schwarzschild AdS black hole for massless and massive particles up to order $\mathcal{O}(G^{2})$. In section \ref{sec:4}, we compute quantum correction to entropy, specific heat and  Helmholtz free energy and pressure. Finally, in section \ref{sec:5}, we summarise our results. We take $c = k_{B} = \bar{h} = 1$ throughout.

\section{Quantum correction to  schwarzschild AdS metric}\label{sec:2}

In this section, we study quantum corrected Schwarzschild AdS metric up to order $\mathcal{O}{(G^2)}$ obtained from Effective Field Theory (EFT). The quantum correction to Schwarzschild metric is given by \cite{calmet2018vanishing}.  The action of the theory is 

\begin{equation}\label{eq:2.1}
    \Gamma[g] = \Gamma_L[g] + \Gamma_{NL}[g].
\end{equation}

where the local part of the action is given by
\begin{equation}\label{eq:2.2}
  \Gamma_L[g] = \int d^4x \sqrt{-g} \Bigl\{  \frac{R + 2 \Lambda}{16 \pi G} + c_1(\mu) R^2 + c_2(\mu) R_{\mu \nu} R^{\mu \nu} + c_3(\mu) R_{\mu \nu \rho \sigma}  R^{\mu \nu \rho \sigma} \Bigl\},  
\end{equation}
where $\Lambda$ is cosmological constant and $\mu$ is an energy scale. The exact values of the coefficients $c_1$, $c_2$ and $c_3$ are unknown, as
they depend on the nature of the ultra-violet theory of quantum gravity. The non-local part of the action is

\begin{equation}\label{eq:2.3}
    \Gamma_{NL}(g) = - \int d^4x \sqrt{-g} \Bigl\{  \alpha R \ln(\frac{\Box}{\mu^2}) R + \beta  R_{\mu \nu} \ln(\frac{\Box}{\mu^2}) R^{\mu \nu} + \gamma R_{\mu \nu \rho \sigma} \ln(\frac{\Box}{\mu^2}) R^{\mu \nu \rho \sigma} \Bigl\},
\end{equation}

the parameters $\alpha$, $\beta$ and $\gamma$ are non-local Wilson coefficients for fields. The values of the parameters for different fields have been listed in ref. \cite{calmet2019quantum}. $\Box = g^{\mu \nu} \nabla_{\mu} \nabla_{\nu} $ and the operator  $\ln( \frac{\Box}{\mu^2})$ has integral representation\cite{donoghue2015covariant}

\begin{equation}\label{eq:2.4}
    \ln(\frac{\Box}{\mu^2}) = \int_{0}^{\infty} ds \Bigl( \frac{1}{\mu^2 +s} - \frac{1}{\Box + s}    \Bigl).
\end{equation}

Using both the local and non-local Gauss-Bonnet identities \cite{calmet2018vanishing}, it is possible to eliminate the Riemann tensor by redefining the coefficients as

\begin{equation}\label{eq:2.5}
\begin{split}
    c_1 \rightarrow \bar{c_1} = c_1 - c_3 , \hspace{1cm} c_2 \rightarrow \bar{c_2} = c_2 + 4c_3 ,\hspace{1cm}   c_3 \rightarrow \bar{c_3} =0, \\ 
    \alpha \rightarrow  \bar{\alpha} = \alpha - \gamma, \hspace{1cm} \beta \rightarrow \bar{\beta} = \beta  + 4 \gamma, \hspace{1cm} \gamma \rightarrow \bar{\gamma} = 0.
    \end{split}
\end{equation}

The quantum corrected Einstein equations can be obtained by varying action \ref{eq:2.1} with respect to the metric,

\begin{equation}\label{eq:2.6}
   \bigl( G_{\mu \nu} - \Lambda g_{\mu \nu} \bigl) + 16 \pi G \bigl( H_{\mu \nu}^{L} + H_{\mu \nu}^{NL} \bigl) =0.
\end{equation}

where $G_{\mu \nu}$ is Einstein tensor given by

\begin{equation}\label{eq:2.7}
    G_{\mu \nu} = R_{\mu \nu} - \frac{1}{2} R g_{\mu \nu},
\end{equation}

The local part of the equation of motion is given by

\begin{equation}\label{eq:2.8}
\begin{split}
    H_{\mu \nu}^{L} = \bar{c_1} \Bigl( 2 R R_{\mu \nu} - \frac{1}{2}g_{\mu \nu}R^2 -2 \nabla_{\mu} \nabla_{\nu}R + 2g_{\mu \nu} \Box R    \Bigl)\\
   + \bar{c_2} \Bigl(  2R^{\rho \sigma} R_{\mu \rho \nu \sigma} -\frac{1}{2}g_{\mu \nu} R_{\rho \sigma}  R^{\rho \sigma} - \nabla_{\mu}\nabla_{\nu}R +\Box R_{\mu \nu} + \frac{1}{2} g_{\mu \nu} \Box R \Bigl),
\end{split}
\end{equation}
The non-local contribution is
\begin{equation*}
          H_{\mu \nu}^{NL} = -2\bar{\alpha} \Bigl(   R_{\mu \nu} - \frac{1}{4}g_{\mu \nu}R + g_{\mu \nu} \Box - \nabla_{\mu} \nabla_{\nu} \Bigl) \ln(\frac{\Box}{\mu^2})R - \bar{\beta} \Bigl( \delta^{\rho}_{\mu}R_{\nu \sigma}  +\delta^{\rho}_{\nu}R_{\mu \sigma} -\frac{1}{2}g_{\mu \nu} R^{\rho}_{\sigma} + \delta^{\rho}_{\mu}g_{\nu \sigma} \Box 
\end{equation*}
\begin{equation}\label{eq:2.9}
     +g_{\mu \nu}\nabla^{\rho}\nabla_{\sigma} - \delta^{\rho}_{\mu} \nabla_{\rho} \nabla_{\nu} - \delta^{\rho}_{\nu} \nabla_{\sigma} \nabla_{\mu} \Bigl) \ln(\frac{\Box}{\mu^2})R^{\sigma}_ {\rho}.
\end{equation}

Classical four dimension Schwarzschild AdS black hole metric is\cite{socolovsky2017schwarzschild}
\begin{equation}\label{eq:2.10}
    ds^2 = \Bigl( 1- \frac{2GM}{r} + \frac{r^2}{a^2}  \Bigl)dt^2 - \Bigl( 1- \frac{2GM}{r} + \frac{r^2}{a^2}  \Bigl)^{-1} dr^2 + r^2d{\Omega}^2,
\end{equation}
where $d{\Omega}^2 = d{\theta}^2 + \sin^2{\theta}d{\phi}^2 $, $a$ is the AdS radius \& it is related to the cosmological constant 
\begin{equation} \label{eq:2.11}
    a=\sqrt{\frac{3}{|\Lambda|}}.
\end{equation}

Now we perturb metric,
\begin{equation}\label{eq:2.12}
    \bar{g}_{\mu \nu} = g_{\mu \nu} + g_{\mu \nu}^{q}.
\end{equation}

We now consider a small perturbation of order $\mathcal{O}(G)$. The equation of motion becomes

\begin{equation}\label{eq:2.13}
 \Bigl(   G_{\mu \nu}^{L}[g^q] - \Lambda g_{\mu \nu} [g^{q}]  \Bigl) + 16 \pi G \Bigl( H_{\mu \nu}^{L}[q] +  H_{\mu \nu}^{NL}[q] \Bigl) =0.
\end{equation}

In case of Schwarzschild AdS solution Ricci scalar and tensor, both are non zero. The action of $\ln(\frac{\Box}{{\mu}^{2}})$ on several radial functions are collected in \cite{calmet2019quantum,campos2022quantum,pourhassan2022quantum}. The final solution of the quantum corrected equations of motion \eqref{eq:2.13} outside $(r > R)$ the star is 

\begin{equation}\label{eq:2.14}
    ds^2 = - f(r)dt^2 + g(r)dr^2 +  r^2d{\Omega}^2,
\end{equation}

where $d{\Omega}^2 = d{\theta}^2 + \sin^2{\theta}d{\phi}^2$ and the metric function is given by

\begin{equation}\label{eq:2.15}
    f(r) =  \Bigl(   1- \frac{2GM}{r} + \frac{r^2}{a^2}     \Bigl) + \tilde{\alpha} \frac{2G^2M}{R^3} \Bigl(  2\frac{R}{r} + \ln(\frac{r-R}{r+R})  \Bigl) + \mathcal{O}(G^3),
\end{equation}

\begin{equation}\label{eq:2.16}
    g(r) =  \Bigl( 1- \frac{2GM}{r} + \frac{r^2}{a^2}  \Bigl)^{-1} +\tilde{\beta} \frac{2G^2M}{r(r^2-R^2)} + \mathcal{O}(G^3),
\end{equation}

with 
\begin{equation}\label{eq:2.17}
\tilde{\alpha} = 96\pi(\alpha + \beta + 3\gamma),
\end{equation}
\begin{equation}\label{eq:2.18}
    \tilde{\beta} = 192\pi(\gamma - \alpha).
\end{equation}
and M is the mass of black hole. The correction of the metric include two parts : one is from local part i.e. eq., \eqref{eq:2.2} \& other is from non-local part i.e.\eqref{eq:2.3}. The correction from local part is trivially 0. Inside the star this is not the case. However, these corrections turn out to be $\mathcal{O}(G^3)$, and thus sub-leading. Therefore the local part in
the equations of motion \eqref{eq:2.8} does not contribute.  The two terms $\tilde{\alpha}$ and $\tilde{\beta}$ are the contributions of non-local part of i.e., eq.\eqref{eq:2.3}.

\section{Quantum correction to Hawking temperature for  Schwarzschild AdS
black hole}\label{sec:3}

In this section, we study the semiclassical tunnelling of particles through the horizon of Schwarzschild black holes, in line with the analysis of \cite{johnson2020hawking}. The
rate of emission $\Gamma$ will have exponential part given by

\begin{equation}  \label{eq:3.1}
    \Gamma \sim \exp(-2ImS),
\end{equation}
where S is the tunnelling action of particles and ImS is the imaginary part of the tunnelling action. According to Plank radiation law, the emitted rate $\Gamma$ of particles with frequency $\omega$ can be written as
\begin{equation} \label{eq:3.2}
    \Gamma \sim \exp(-\frac{\omega}{T_{BH}}).
\end{equation}

Hence the temperature at which black hole radiates given by
\begin{equation} \label{eq:3.3}
    T_{BH} = \frac{\omega}{2ImS}.
\end{equation}

In this section,  we will study the process in Painlevé-Gullstrand coordinates. Firstly, we need to rewrite the corrected metric in Painlevé-Gullstrand coordinates to order $G^2$ We define
\begin{equation} \label{eq:3.4}
    t_{r} =t-a(r),
\end{equation}

where $a(r)$ is function of r. Than 
\begin{equation} \label{eq:3.5}
    dt^2 = dt_r^2+ a^{\prime 2}(r) dr^2 + 2a^{\prime}(r)dt_rdr,
\end{equation}

where $a^{\prime}(r)= \frac{da(r)}{dr}$. Using equation \eqref{eq:3.5} \& \eqref{eq:2.14} we have

\begin{equation} \label{eq:3.6}
    ds^2= -f(r)dt_{r}^2 +\{ g(r) -f(r) a^{\prime 2}(r) \} dr^2 -2a^{\prime}(r) f(r) dt_{r}dr +r^2 d{\Omega}^{2}.
\end{equation}

We require
\begin{equation} \label{eq:3.7}
   \{ g(r) -f(r) a^{\prime 2}(r) \} =1, 
\end{equation}

we have $a^{\prime 2} = \frac{g(r) - 1}{f(r)}$. We take $a^{\prime} =- \sqrt{\frac{g(r) - 1}{f(r)}}$. Than

\begin{equation} \label{eq:3.8}
    -2a^{\prime}f(r) = 2 \sqrt{f(r) \{g(r) - 1\}},
\end{equation}

For classical Schwarzschild AdS metric, $ f(r) = (1- \frac{2GM}{r} + \frac{r^2}{a^2})$ and $ g(r) = (1- \frac{2GM}{r} + \frac{r^2}{a^2})^{-1}$, we have

\begin{equation} \label{eq:3.9}
    ds^2 = -(1- \frac{2GM}{r} + \frac{r^2}{a^2}) dt_{r}^2 + dr^2 +2\sqrt{(1+ \frac{r^2}{a^2})\frac{2GM}{r}-(1- \frac{2GM}{r} + \frac{r^2}{a^2}) \frac{r^2}{a^2}} \hspace{1mm} dt_{r} dt + r^2 d{\Omega}^2.
\end{equation}

For the quantum corrected Schwarzschild AdS metric, f(r) and g(r) have been given in equations \eqref{eq:2.15} and \eqref{eq:2.16}, then we have

\begin{equation} \label{eq:3.10}
    f(r) (g(r) -1) = (1+ \frac{r^2}{a^2})\frac{2GM}{r}-(1- \frac{2GM}{r} + \frac{r^2}{a^2}) \frac{r^2}{a^2} + (1+ \frac{r^2}{a^2}) \beta -\frac{r^2}{a^2} \alpha,
\end{equation}

\begin{equation}\label{eq:3.11}
    -2a^{\prime}f(r) = 2 \sqrt{f(r) \{g(r) - 1\}}=  2 \sqrt{(1+ \frac{r^2}{a^2})\frac{2GM}{r}-(1- \frac{2GM}{r} + \frac{r^2}{a^2}) \frac{r^2}{a^2} + (1+ \frac{r^2}{a^2}) \beta -\frac{r^2}{a^2} \alpha},
\end{equation}

Therefore, to order $\mathcal{O}(G^2)$ the quantum corrected Schwarzschild AdS metric can be written as, 

\begin{equation}\label{eq:3.12}
    ds^2= -f(r)dt_{r}^2 + dr^2 +2 \sqrt{(1+ \frac{r^2}{a^2})\frac{2GM}{r}-(1- \frac{2GM}{r} + \frac{r^2}{a^2}) \frac{r^2}{a^2} + (1+ \frac{r^2}{a^2}) \beta -\frac{r^2}{a^2} \alpha} \hspace{3mm}dt_{r}dr +r^2 d{\Omega}^{2},
\end{equation}

Dropping the  subscript r for convenience, i.e.

\begin{equation}\label{eq:3.13}
    ds^2= -f(r)dt^2 + dr^2 +2 \sqrt{(1+ \frac{r^2}{a^2})\frac{2GM}{r}-(1- \frac{2GM}{r} + \frac{r^2}{a^2}) \frac{r^2}{a^2} + (1+ \frac{r^2}{a^2}) \beta -\frac{r^2}{a^2} \alpha} \hspace{3mm}dt dr +r^2 d{\Omega}^{2}.
\end{equation}

The action for a particle moving freely in a curved background can be written as

\begin{equation}\label{eq:3.14}
    S= \int p_{\mu} dx^{\mu}, 
\end{equation}

where,
\begin{equation}\label{eq:3.15}
p_{\mu} =m g_{\mu \nu} \frac{dx^{\nu}}{d\sigma}.
\end{equation}

where $\sigma$ is an affine parameter along the worldline of the particle, chosen so that $p_{\mu}$ coincides with the physical 4-momentum of the particle. For a massive particle, this requires that $ d\sigma = \frac{d\tau}{m}$, with $\tau$ is the proper time. 

\subsection{Quantum correction to Hawking temperature}

\subsubsection{Massless particles}

The radial dynamics of massless particles in Schwarzschild spacetime are determined by the equations 

\begin{equation}\label{eq:3.16}
    \dot{r}^{2} + 2 \sqrt{(1+ \frac{r^2}{a^2})\frac{2GM}{r} + (1+ \frac{r^2}{a^2}) \beta -\frac{r^2}{a^2} f(r)} \hspace{2mm} \dot{t} \dot{x} - f(r) \dot{t}^{2}=0.
\end{equation}

From equation \eqref{eq:3.15} taking $\mu=t$ 

\begin{equation}\label{eq:3.17}
    p_{t} = g_{tt} \dot{t} + g_{tr} \dot{r},
\end{equation}

$p_{t}=- \omega$ and $\omega$  has the interpretation of the energy of the particle as measured at infinity. Equation \eqref{eq:3.16} can be factorised to the equation:

\begin{equation}\label{eq:3.18}
    \dot{r} = \biggr[\pm( 1-\sqrt{\frac{2GM}{r}})(1+ \frac{r^2}{2a^{2}}) - \sqrt{\frac{2GM}{r}} \frac{\tilde{\beta}G}{2(r^{2}-R^{2})}(1+\frac{r^2}{a^{2}})   \biggr] \dot{t},
\end{equation}

with the positive sign for outgoing and negative sign for ingoing geodesics. We only consider the case of outgoing particles. From equations \eqref{eq:3.17} and \eqref{eq:3.18} we have

\begin{equation}\label{eq:3.19}
    \dot{t} = \frac{\omega}{1-\sqrt{\frac{2GM}{r}}(1+\frac{r^{2}}{a^{2}}) - \sqrt{\frac{2GM}{r}}\frac{\tilde{\beta}G}{2(r^{2}-R^{2})}(1+\frac{r^{2}}{a^{2}})},
\end{equation}

From equation  \eqref{eq:3.14}, \eqref{eq:3.15} \& \eqref{eq:3.19} we have 

\begin{equation}\label{eq:3.20}
  ImS = Im \int p_{r} dr  =  Im \int \dot{t} dr.
\end{equation}

Define

\begin{equation}\label{eq:3.21}
  1-\sqrt{\frac{2G\tilde{M}}{r}}  = 1-\sqrt{\frac{2GM}{r}}(1+\frac{r^{2}}{a^{2}}) - \sqrt{\frac{2GM}{r}}\frac{\tilde{\beta}G}{2(r^{2}-R^{2})}(1+\frac{r^{2}}{a^{2}})
\end{equation},

Than equation  \eqref{eq:3.20} becomes 

\begin{equation}\label{eq:3.22}
     ImS = Im \int \dot{t} dr= \int\frac{\omega dr}{1- \sqrt{\frac{2G\tilde{M}}{r}}}.
\end{equation}

The integrand has a pole at $r = 2\tilde{M} $, the horizon. Choosing the prescription to integrate clockwise around this pole (into the upper-half complex-r plane) \cite{johnson2020hawking}, \cite{liu2022effect} we find

\begin{equation}\label{eq:3.23}
     ImS = 4 \pi G \omega \tilde{M}.
\end{equation}

Than the temperature at which the black hole radiates, equation \eqref{eq:3.3} becomes 

\begin{equation}\label{eq:3.24}
      T_{BH} = \frac{\omega}{2ImS}=\frac{1}{8 \pi G \tilde{M}}.
\end{equation}

From equation \eqref{eq:3.21}

\begin{equation}\label{eq:3.25}
    \tilde{M}=M \biggr[ 1+ \frac{\tilde{\beta}G}{2(r^{2}-R^{2})}  \biggr]^{2} \Bigl(1+\frac{r^2}{a^2}\Bigl)^{2}.
\end{equation}

Therefor 

\begin{equation}\label{eq:3.26}
    T_{BH} = \frac{1}{8 \pi G M} \biggl( 1- \frac{\tilde{\beta}G}{(r^{2} - R^{2})}- \frac{\tilde{\beta}^{2}G^{2}}{4(r^{2} - R^{2})^2} \biggl) \biggl( 1 - 2 \frac{r^{2}}{a^{2}} - \frac{r^{4}}{a^{4}} \biggl) .
\end{equation}

\subsubsection{Massive particles}

For massive point-particles, the dynamics are governed by

\begin{equation}\label{eq:3.27}
    \dot{r}^{2} + 2 \sqrt{(1+ \frac{r^2}{a^2})\frac{2GM}{r} + (1+ \frac{r^2}{a^2}) \beta -\frac{r^2}{a^2} f(r)} \hspace{2mm} \dot{t} \dot{r} +(1- f(r) \dot{t}^{2})=0.
\end{equation}

From equation  \eqref{eq:3.15} taking $\mu=t$ 

\begin{equation}\label{eq:3.28}
    p_{t} = g_{tt} \dot{t} + g_{tr} \dot{r}.
\end{equation}

$p_{t}= -\omega$ and $\omega$  has the interpretation of the energy of the particle as measured at infinity. Equation  \eqref{eq:3.27}  can be factorised to the equation

\begin{equation}\label{eq:3.29}
    \dot{r} = \biggr[\pm( 1-\sqrt{\frac{2GM}{r}})(1+ \frac{r^2}{2a^{2}}) - \sqrt{\frac{2GM}{r}} \frac{\tilde{\beta}G}{2(r^{2}-R^{2})}(1+\frac{r^2}{a^{2}})   \biggr] \dot{t},
\end{equation}

with the positive sign for outgoing and negative sign for ingoing geodesics. We only consider the case of outgoing particles. From equations \eqref{eq:3.28} \& \eqref{eq:3.29} we have

\begin{equation}\label{eq:3.30}
    \dot{t} = \frac{\omega}{1-\sqrt{\frac{2GM}{r}}(1+\frac{r^{2}}{a^{2}}) - \sqrt{\frac{2GM}{r}}\frac{\tilde{\beta}G}{2(r^{2}-R^{2})}(1+\frac{r^{2}}{a^{2}})},
\end{equation}

From equation  \eqref{eq:3.14},  \eqref{eq:3.15} \& \eqref{eq:3.30}, we have 

\begin{equation}\label{eq:3.31}
  ImS = Im \int p_{r} dr  =  Im \int \dot{t} dr.
\end{equation}

Define

\begin{equation}\label{eq:3.32}
  1-\sqrt{\frac{2G\tilde{M}}{r}}  = 1-\sqrt{\frac{2GM}{r}}(1+\frac{r^{2}}{a^{2}}) - \sqrt{\frac{2GM}{r}}\frac{\tilde{\beta}G}{2(r^{2}-R^{2})}(1+\frac{r^{2}}{a^{2}}),
\end{equation}

Than equation  \eqref{eq:3.31} becomes 

\begin{equation}\label{eq:3.33}
     ImS = Im \int \dot{t} dr= \int\frac{\omega dr}{1- \sqrt{\frac{2G\tilde{M}}{r}}}.
\end{equation}

The integrand has a pole at $r = 2\tilde{M} $, the horizon. Choosing the prescription to integrate clockwise around this pole (into the upper-half complex-r plane) \cite{johnson2020hawking}, \cite{liu2022effect} we find

\begin{equation}\label{eq:3.34}
     ImS = 4 \pi G \omega \tilde{M}.
\end{equation}

Than the temperature at which the black hole radiates, equation \eqref{eq:3.3} becomes 

\begin{equation}\label{eq:3.35}
      T_{BH} = \frac{\omega}{2ImS}=\frac{1}{8 \pi G \tilde{M}}.
\end{equation}

From equation \eqref{eq:3.32}

\begin{equation}\label{eq:3.36}
    \tilde{M}=M \biggr[ 1+ \frac{\tilde{\beta}G}{2(r^{2}-R^{2})}  \biggr]^{2} \Bigl(1+\frac{r^2}{a^2}\Bigl)^{2},
\end{equation}

Therefor 

\begin{equation}\label{eq:3.37}
    T_{BH} = \frac{1}{8 \pi G M} \biggl( 1- \frac{\tilde{\beta}G}{(r^{2} - R^{2})}- \frac{\tilde{\beta}^{2}G^{2}}{4(r^{2} - R^{2})^2} \biggl) \biggl( 1- 2 \frac{r^{2}}{a^{2}}- \frac{r^{4}}{a^{4}} \biggl). 
\end{equation}
Same as equation \eqref{eq:3.26}.

\section{Quantum correction to thermodynamics of Schwarzschild
AdS black hole}\label{sec:4}
In this section we study the quantum corrected black hole entropy and various thermodynamics parameters.  We find entropy of Schwarzschild
AdS black hole upto order $\mathcal{O}{(G^{2})}$ and plot the classical \& quantum entropy. Next we study specific heat and its comes out positive. Finally we compute Helmholtz free energy and pressure of quantum corrected Schwarzschild AdS black hole and we plot it's classical and quantum version.

\subsection{Quantum correction to entropy}
The entropy of the system is defined as
\begin{equation}\label{eq:4.1}
    S=\frac{\pi {r_{h}}^{2}}{G}= \frac{\pi {(2G \tilde{M})}^{2}}{G},
\end{equation}
\begin{equation}\label{eq:4.2}
    S= 4{\pi}GM^{2}  \biggr[ 1+ \frac{\tilde{\beta}G}{(r^{2}-R^{2})} +\frac{\tilde{\beta}^{2}G^{2}}{4(r^{2}-R^{2})^{2}} \biggr]^{2} \Bigl(1+\frac{2r^2}{a^2} + \frac{r^4}{a^4}\Bigl)^{2},
\end{equation}
Now substituting $M$ from equation \eqref{eq:3.37} and taking term upto order $\mathcal{O}{(G^{2})}$ we have 
\begin{equation}\label{eq:4.3}
    S= \frac{1}{16 \pi T_{BH}^{2}}  \biggr[ 1+ \frac{2\tilde{\beta}G}{(r^{2}-R^{2})} +\frac{3\tilde{\beta}^{2}G^{2}}{2(r^{2}-R^{2})^{2}} \biggr] \biggr[ 1- \frac{2\tilde{\beta}G}{(r^{2}-R^{2})} +\frac{\tilde{\beta}^{2}G^{2}}{2(r^{2}-R^{2})^{2}} \biggr] \Bigl(1+\frac{2r^2}{a^2} + \frac{r^4}{a^4}\Bigl)^{2} \Bigl(1-\frac{2r^2}{a^2} - \frac{r^4}{a^4}\Bigl)^{2},
\end{equation}
\begin{equation}\label{eq:4.4}
    S=\frac{1}{16 \pi T_{BH}^{2}}  \biggr[ 1 - \frac{2\tilde{\beta}^{2} G^{2}}{(r^2 - R^2)^{2}}\biggr] \Bigl(1+\frac{2r^2}{a^2} + \frac{r^4}{a^4}\Bigl)^{2} \Bigl(1-\frac{2r^2}{a^2} - \frac{r^4}{a^4}\Bigl)^{2}.
\end{equation}

\begin{figure}[H]
\centering
\subfloat[]{\includegraphics[width=.5\textwidth]{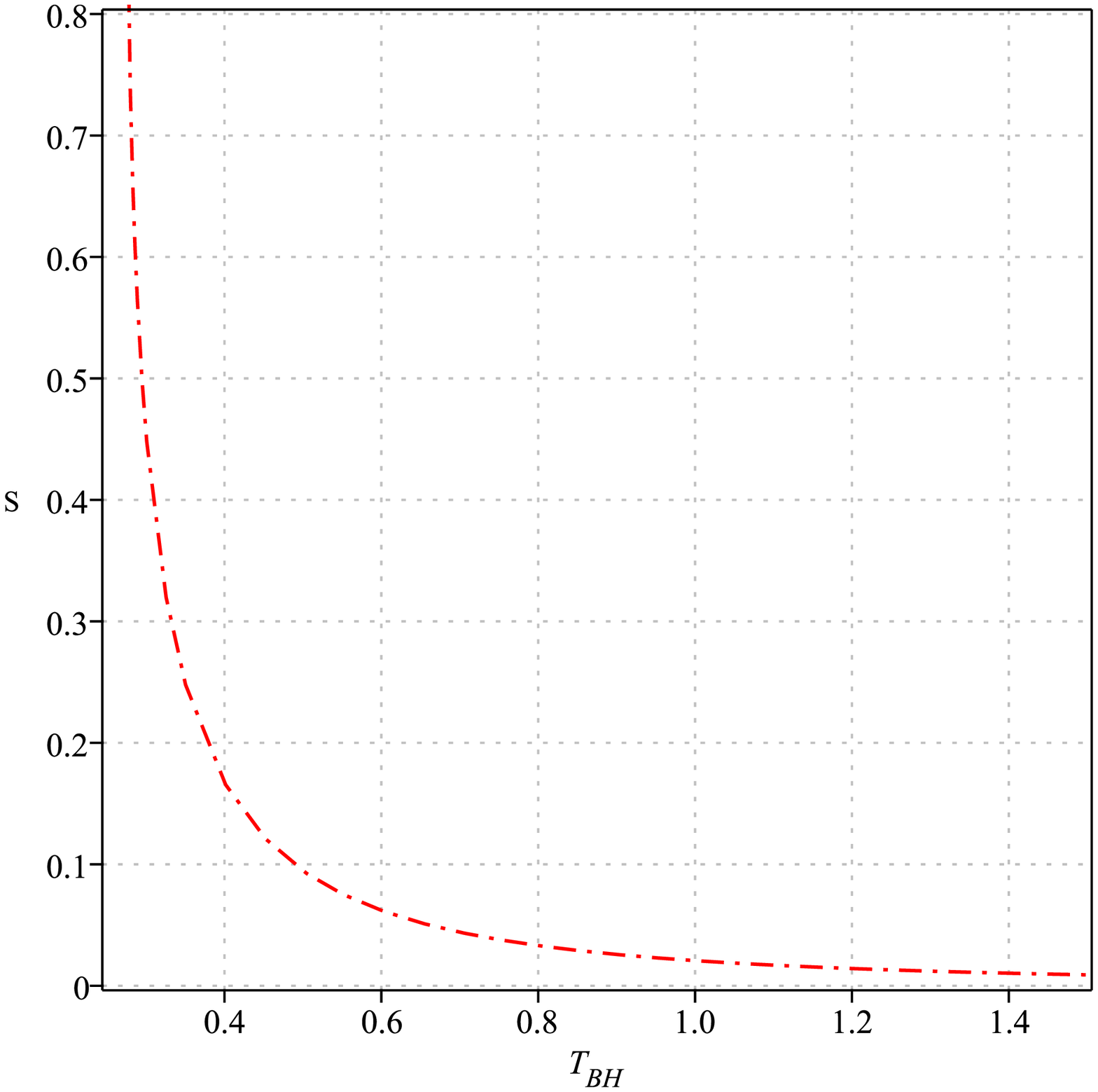}}\hfill
\subfloat[]{\includegraphics[width=.5\textwidth]{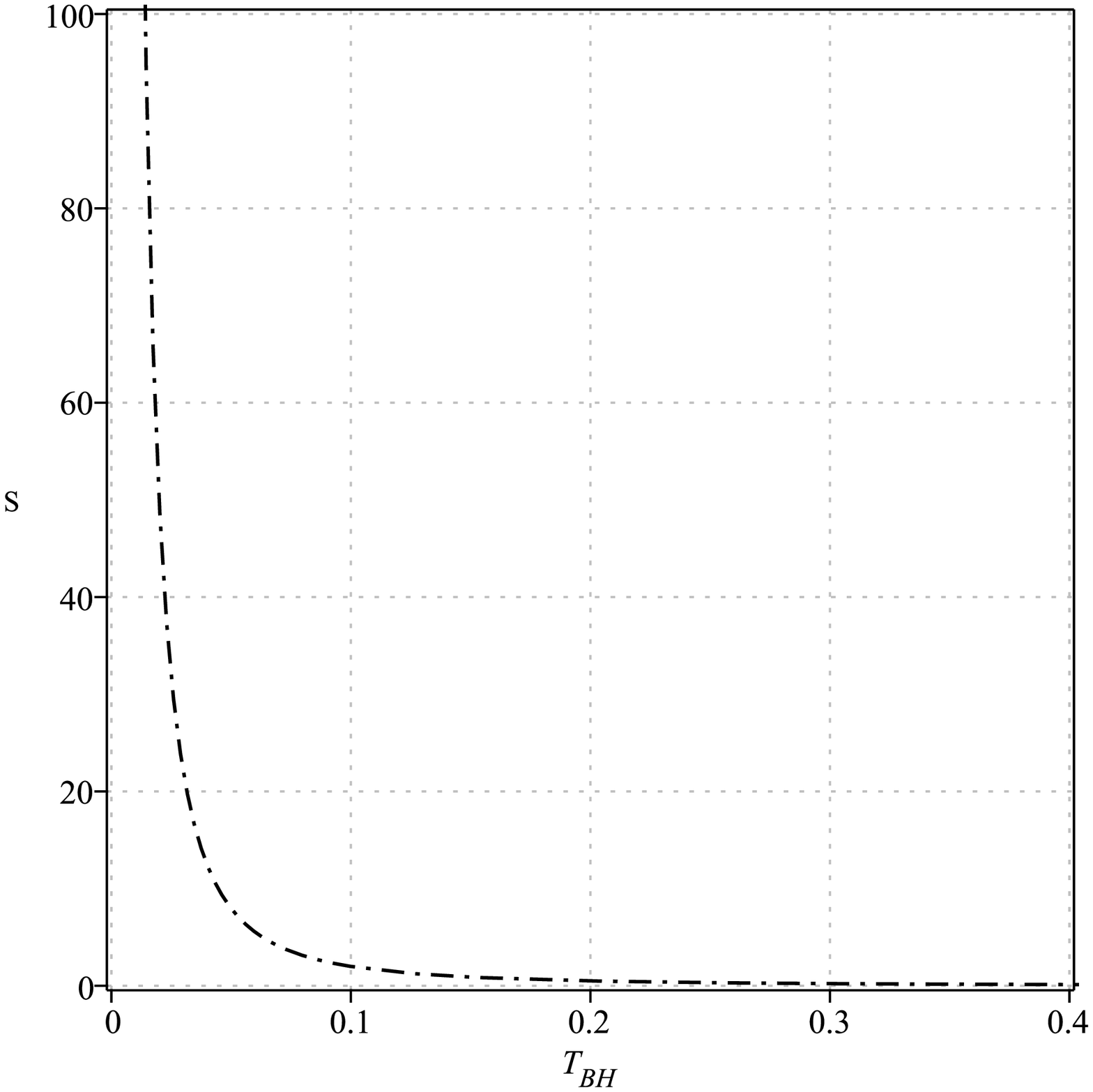}}\hfill
\caption{}\label{fig:1}
\end{figure}
\textbf{Figure: (a)} Entropy of Classical Schwarzschild AdS black hole with $G=1$, $r=3R$,  $M=2$ and $a=1.5$.  \textbf{Figure: (b)} Entropy of quantum corrected Schwarzschild AdS black hole with $G=1$, $r=3R$,  $M=2$ and $a=1.5$.  Numerical values of $\alpha$, $\beta$ and $\gamma$ taken from  \cite{calmet2019quantum}.  All fields (Scalar, Fermionic, Vector and Gravitational) give rise to approximately the same behaviour. 

The specific heat of black hole is 
\begin{equation}\label{eq:4.5}
    C=\frac{M}{T_{BH}} \biggl(\frac{dS}{dT_{BH}}\biggl),
\end{equation}
\begin{equation}\label{eq:4.6}
    C= \frac{G \biggr[ 1+ \frac{2\tilde{\beta}G}{(r^{2}-R^{2})} +\frac{3\tilde{\beta}^{2}G^{2}}{2(r^{2}-R^{2})^{2}} \biggr] \biggr[ 1- \frac{2\tilde{\beta}G}{(r^{2}-R^{2})} +\frac{\tilde{\beta}^{2}G^{2}}{2(r^{2}-R^{2})^{2}} \biggr] \Bigl(1+\frac{2r^2}{a^2} + \frac{r^4}{a^4}\Bigl)^{2} \Bigl(\frac{2r^2}{a^2} + \frac{r^4}{a^4} - 1\Bigl)}{T_{BH} \biggr[ 1-\frac{\tilde{\beta} G}{(r^{2} - R^{2})} - \frac{\tilde{\beta}^{2} G^{2}}{4(r^{2} - R^{2})^{2}}  \biggr] }.
\end{equation}

Equations \eqref{eq:3.36}, \eqref{eq:3.37}, \eqref{eq:4.2} and \eqref{eq:4.6} depends on $R$. Since the quantum modification are tiny as the term $\frac{\tilde{\beta} G}{(r^2 - R^2)}$ is very small. Classical entropy $(S=\frac{A}{4G})$ of black hole get modified, as there is a shift in the position of horizon.

\subsection{Quantum correction to free energy and pressure}
The Helmholtz free energy is
\begin{equation}\label{eq:4.7}
    F= \frac{GM \biggr[ 1+ \frac{2\tilde{\beta}G}{(r^{2}-R^{2})} +\frac{3\tilde{\beta}^{2}G^{2}}{2(r^{2}-R^{2})^{2}} \biggr] \biggr[ 1- \frac{2\tilde{\beta}G}{(r^{2}-R^{2})} +\frac{\tilde{\beta}^{2}G^{2}}{2(r^{2}-R^{2})^{2}} \biggr] \Bigl(1+\frac{2r^2}{a^2} + \frac{r^4}{a^4}\Bigl)^{2} \Bigl(\frac{2r^2}{a^2} + \frac{r^4}{a^4} -1\Bigl)}{2 \biggr[  1- \frac{\tilde{\beta}G}{(r^{2} - R^{2})}- \frac{\tilde{\beta}^{2}G^{2}}{4(r^{2} - R^{2})^2}  \biggr]} +M .
\end{equation}

Quantum effects also produce modification on the pressure. For a schwarzschild AdS black hole pressure is 

\begin{equation}\label{eq:4.8}
    P = - \frac{T_{BH} \frac{dS}{dM} - 1}{\frac{dV}{dM}}
\end{equation}

\begin{equation}\label{eq:4.9}
    P= \frac{\biggr[ 1- \frac{\tilde{\beta}G}{(r^{2} - R^{2})}- \frac{\tilde{\beta}^{2}G^{2}}{4(r^{2} - R^{2})^2}  \biggr] \biggr[  1+ \frac{\tilde{\beta}G}{(r^{2} - R^{2})}+ \frac{\tilde{\beta}^{2}G^{2}}{4(r^{2} - R^{2})^2} \biggr]^{2} \biggl(  2 \frac{r^{2}}{a^{2}}+ \frac{r^{4}}{a^{4}} -1\biggl)  \biggl( 1+ 2 \frac{r^{2}}{a^{2}}+ \frac{r^{4}}{a^{4}} \biggl)^{2} - 1}{32 \pi G^{3} M^{2} \biggr[ 1+ \frac{2 \tilde{\beta} G}{2(r^2 - R^2 )^{2}}  \biggr]^{6} \biggl( 1+ 2 \frac{r^{2}}{a^{2}}+ \frac{r^{4}}{a^{4}} \biggl)^{4}}.
\end{equation}

Next, we will discuss the variation of Hawking temperature for classical and quantum corrected schwarzschild AdS black holes.
\begin{figure}[H]
\centering
\subfloat[]{\includegraphics[width=.5\textwidth]{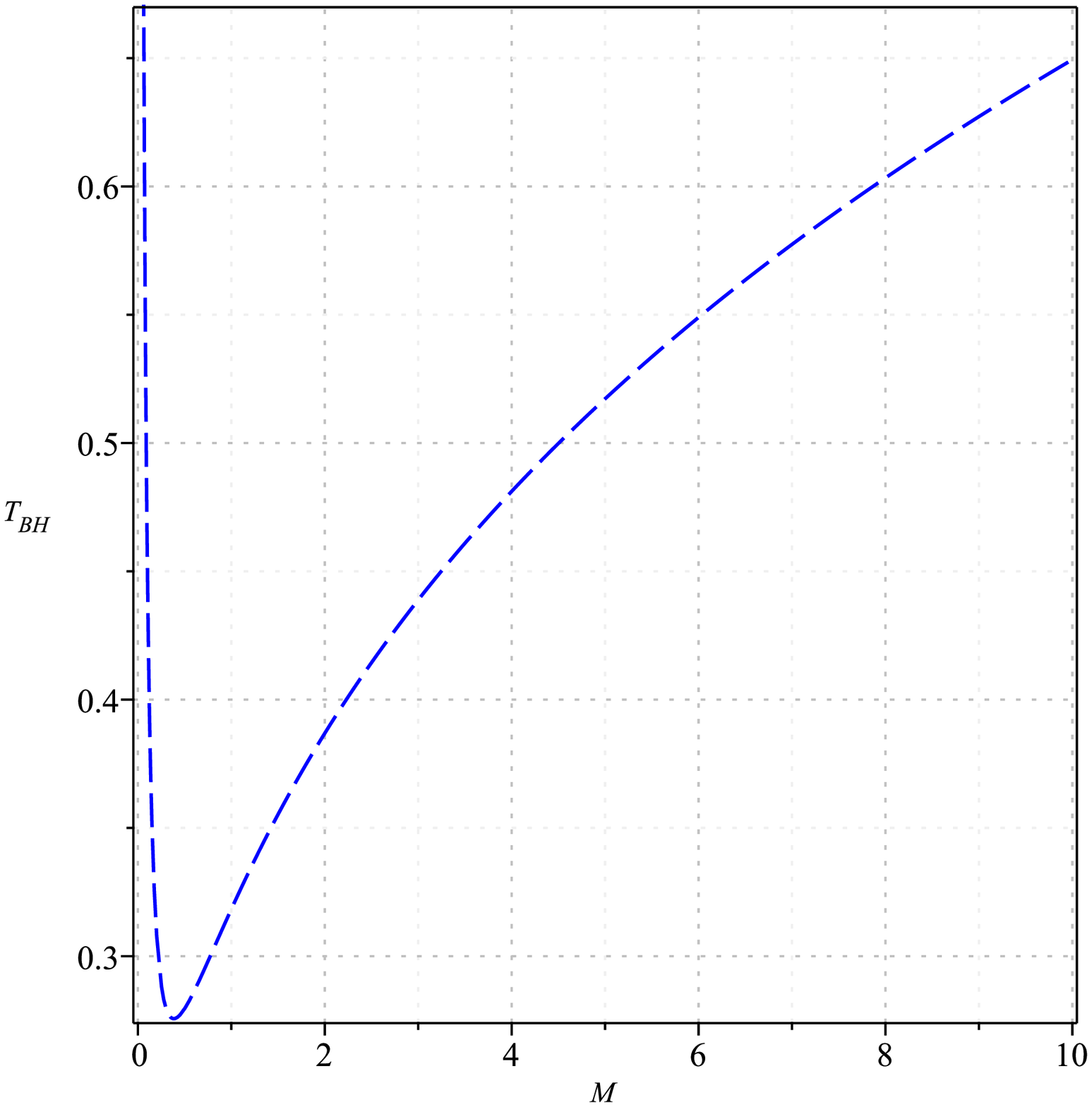}}\hfill
\subfloat[]{\includegraphics[width=.5\textwidth]{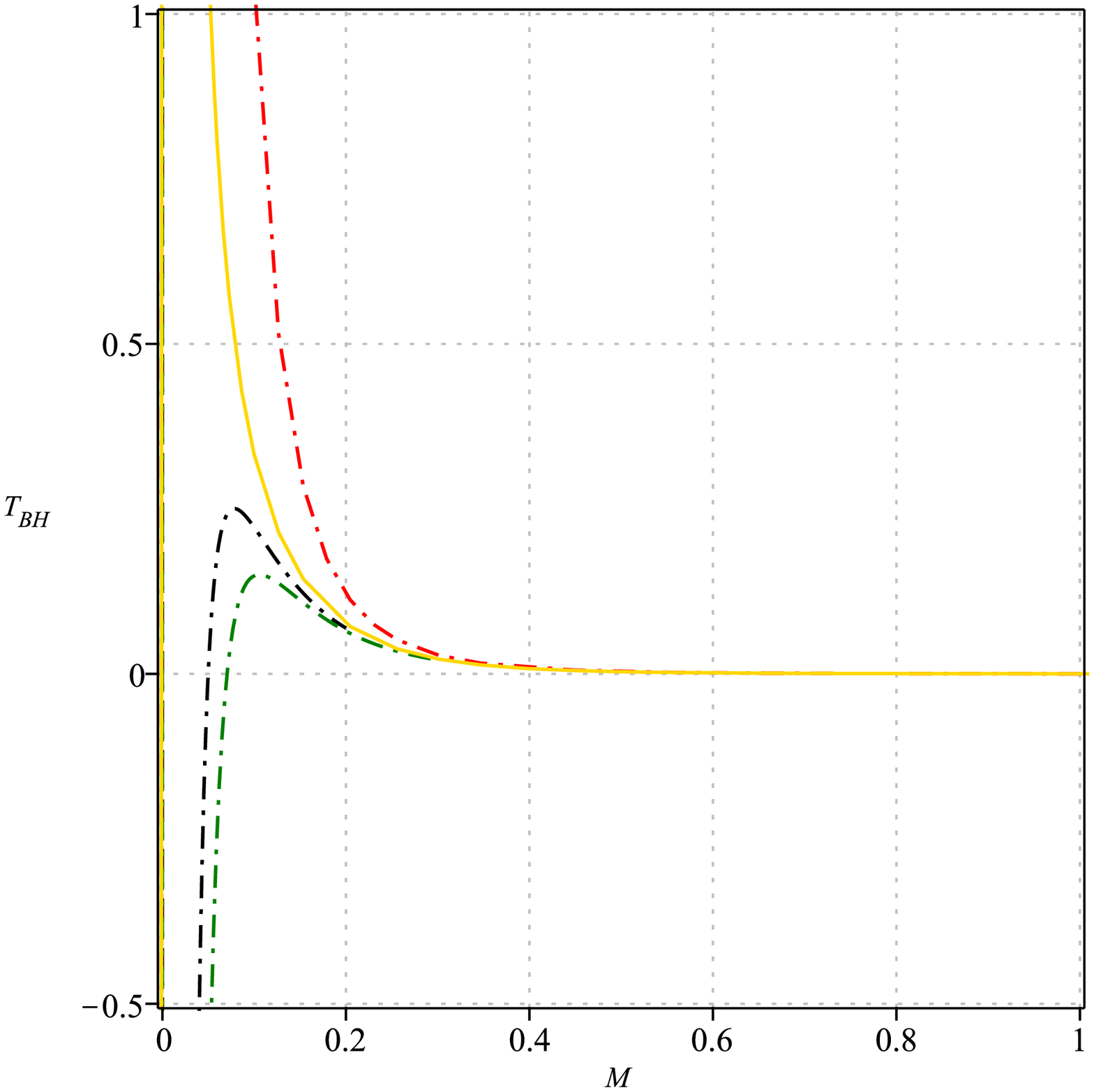}}\hfill
\caption{}\label{fig:2}
\end{figure}

\textbf{Figure: (a) }  Hawking temperature of Classical Schwarzschild AdS black hole with $G=1$, $r=3R$,  $M=2$ and $a=1.5$.  \textbf{Figure: (b) }  Hawking temperature  of quantum corrected Schwarzschild AdS black hole with $G=1$, $r=3R$,  $M=2$ and $a=1.5$.  Dash dot red: Quantum corrected Schwarzschild AdS with scalar field ($ \alpha = 0.0010994$, $\beta = - 0.00001759$ , $ \gamma = 0.00001759 $).  Dash dot black: Quantum corrected Schwarzschild AdS with fermionic field ($ \alpha =-0.000043976 $, $\beta = 0.000070361$, $\gamma =0.000061566 $). Solid green: Quantum corrected Schwarzschild AdS with vector field ( $ \alpha = -0.000433976 $, $\beta = 0.0015480 $, $\gamma = -0.00022867 $).  Solid gold: Quantum corrected Schwarzschild AdS with gravition field ( $ \alpha = 0.0037819 $, $\beta = -0.012700 $, $\gamma = 0.0037292 $).   Numerical values of $\alpha$, $\beta$ and $\gamma$ taken from  \cite{calmet2019quantum}.

\section{Conclusions}\label{sec:5}
In this paper, we consider four-dimensional Anti de Sitter Schwarzschild black holes. In ref. \cite{calmet2019quantum} the authors used EFT methods to obtain the metric inside and outside a static and spherically symmetric star, taking into account quantum effects.  Authors also proposed that the exterior metric may provide a good model of quantum black holes. In section 2, we used EFT method to obtain quantum correction to the four-dimensional Anti de Sitter Schwarzschild  black holes. In section 3, we studied quantum correction of Hawking temperature for massless and massive particle. First we studied the radial dynamic equations of massless and massive particle  in quantum corrected Schwarzschild AdS spacetime. Then, following the spirit of ref.\cite{johnson2020hawking,liu2022effect}, we obtained the modified Hawking temperature. By computing the modifed Hawking entropy, we found that the famous area law of black holes $S = \frac{A}{4G}$ need to be corrected. In section 4, we compute the quantum correction to entropy, specific heat, free energy and pressure.

\printbibliography
\end{document}